\documentclass[a4paper]{article}

\usepackage{INTERSPEECH2022}
\usepackage{tipa}
\usepackage{multirow}
\usepackage{booktabs}
\usepackage{xcolor}

\title{Improving Hypernasality Estimation with Automatic Speech Recognition in Cleft Palate Speech}
\name{
    Kaitao Song$^1$, Teng Wan$^2$, Bixia Wang$^2$, Huiqiang Jiang$^1$,  Luna Qiu$^1$, Jiahang Xu$^1$, \\
    Liping Jiang$^2$, Qun Lou$^2$, Yuqing Yang$^1$, Dongsheng Li$^1$, Xudong Wang$^{2,*}$~\thanks{* Corresponding Author: Xudong Wang}, Lili Qiu$^1$
}
\address{
    $^1$Microsoft Research \\
    $^2$ {Department of Oral and Craniomaxillofacial Surgery, Shanghai Ninth People’s Hospital, \\Shanghai Jiao Tong University School of Medicine}
}
\email{\{kaitaosong, hjiang, lunaqiu, jiahangxu, Yuqing.Yang, dongsli, liliqiu\}@microsoft.com, wonton1984@gmail.com, \{bixia1987, wangxudong70\}@hotmail.com, jlping64@126.com, wonderful{\_}0322@163.com}
\newcommand{\eg}{\emph{e.g.}}
\newcommand{\ie}{\emph{i.e.}}

\begin{document}

\maketitle
\begin{abstract}
Hypernasality is an abnormal resonance in human speech production, especially in patients with craniofacial anomalies such as cleft palate. In clinical application, hypernasality estimation is crucial in cleft palate diagnosis, as its results determine the subsequent surgery and additional speech therapy.  Therefore, designing an automatic hypernasality assessment method will facilitate speech-language pathologists to make precise diagnoses. Existing methods for hypernasality estimation only conduct acoustic analysis based on low-resource cleft palate dataset, by using statistical or neural network-based features. In this paper, we propose a novel approach that uses automatic speech recognition model to improve hypernasality estimation. Specifically, we first pre-train an encoder-decoder framework in an automatic speech recognition (ASR) objective by using speech-to-text dataset, and then fine-tune ASR encoder on the cleft palate dataset for hypernasality estimation. Benefiting from such design, our model for hypernasality estimation can enjoy the advantages of ASR model: 1) compared with low-resource cleft palate dataset, the ASR task usually includes large-scale speech data in the general domain, which enables better model generalization; 2) the text annotations in ASR dataset guide model to extract better acoustic features. Experimental results on two cleft palate datasets demonstrate that our method achieves superior performance compared with previous approaches.

\end{abstract}
\noindent\textbf{Index Terms}: Cleft Palate, Hypernasality, Automatic Speech Recognition.

\section{Introduction}
Cleft lip and palate (CLP)~\cite{Kummer2016Evaluation}, is the most common congenital deformity in the oral and maxillofacial region, with the incidence of 1.4 in 1000 live births in human. In CLP patients, they usually have inability to pronounce the normal sound due to the incomplete closure of their soft palate (\ie, velopharyngeal dysfunction), and thus results in the production of hypernasality~\cite{Vijayalakshmi2007Acoustic,Guo2006Voice}. To treat this illness, CLP palates are usually required to conduct a series of palate surgery and subsequent speech therapy. Therefore, how to estimate the CLP severity is a critical element to determine the final treatment for patients.

Hypernasality is regarded as one of the primary symptoms in cleft palate and it is usually characterized as excessive nasal resonance in pronouciation. As a result, the CLP patients with hypernasality cannot pronounce vowel and consonant as clear as normal people. Therefore, to a certain degree, the severity of hypernasality reflects the degree of the opening and closing of a velopharyngeal passageway between the oral cavity and nasal cavity. How to estimate the severity of hypernasality has been considered as an important metric to evaluate the outcome  of primary cleft palate repair, and to determine the need for the further treatment, such as pharyngoplasty and speech therapy. In clinical examinations, hypernasality rating is usually evaluated by speech-language pathologists (SLPs). However, availability of expert SLPs is usually limited, and the diversity of intra-rater and inter-rater reliablities will also affect the result of subjective hypernasality evaluation. Accordingly, some scientists attempted to investigate the possibility of detecting hypernasality by using machines or instruments. For example, nasometer~\cite{Bettens2014Instrumental} is a widely used instrument to measure nasalance with a rating of 0-100. Nevertheless, these instrumental methods do not demonstrate high correlation with clinical perception of hypernasality and these instruments also require experienced clinicians to operate~\cite{WATTERSON199313}.

Recently, machine learning based automatic hypernasality estimation has drawn enormous attention. Existing methods are mainly based on speech signal processing as cleft palate speech exhibits abnormal nasal resonance in the spectrum features. The training pipeline of machine learning based methods is to first extract acoustic features, like Mel-frequency cepstral coefficients (MFCC) or filter banks, and then perform classification using support vector machine and Gaussian mixture models. For example, \cite{Maier2009Automatic} applied voice intonation features, MFCC and Teager energy operator to detect hypernasality. \cite{He2014Automatic} investigated the effect of energy shifts to the low-frequency bands to estimate hypernasality.  \cite{Mohammad2020Single} introduced a single frequency filter bank based long-term average spectral for hypernasality estimation. \cite{Golabbakhsh2017acoustic} analyzed multiple different acoustic features for automatic identification  of hypernasality. \cite{Dubey2019Hypernasality,Saxon2019Objective} also designed some novel acoustic features to better extract hypernasality-related semantics. These works mainly focused on extracting or designing advanced acoustic features  for hypernasal speech detection. Inspired by the success of deep neural network in speech processing tasks, some works~\cite{Vikram2021Attention,Xiyue2019HypernasalityNet,Vikram2021Deep,Wang2019cnn} also tried to use recurrent neural network, convolutional network or attention model to detect hypernasality in an end-to-end manner.

Although significant progresses have been achieved in evaluating hypernasality, there still remain a few limitations in the current systems. First, existing systems for hypernasality estimation only leverage cleft palate datasets. However,  different from common speech processing tasks, cleft palate datasets are usually extremely difficult to collect, due to privacy and its annotations that require expert SLPs labeling. In other words, CLP datasets are usually in a low-resource scenario, which results in a poor generalization for model capacity. Secondly, previous work~\cite{bechet12Consonantal,Nikitha2017Vowel} have analyzed that hypernasality has obviously abnormal area in acoustic space (\eg, vowel or consonant), especially in the spectrum dimension~\cite{Xiyue2019HypernasalityNet}. Thus, it can be concluded that learning high quality acoustic features is beneficial for predicting hypernasality. However, existing systems (especially neural network based models) for hypernasality estimation only build the connection between the CLP speech and its hypernasality rating, without modeling acoustic features explicitly or implicitly. How to address the above two items for hypernasality estimation is the target of our paper. 


Different from CLP datasets, many speech processing tasks in general domain usually have rich resource. Specifically, automatic speech recognition (ASR)~\cite{James1990ASR} is one of the most representative tasks in speech processing. The objective of ASR is to recognize the content of the input speech and return its corresponding text. By leveraging additional text annotations, the model is able to understand the content of speech semantic (\eg, vowel or consonant) and extract high quality representations for acoustic features. In addition, many ASR datasets~\cite{Vassil2015Librispeech,Hui2017Aishell} usually have enormous speech data (\eg, Librispeech~\cite{Vassil2015Librispeech} includes 960 hrs data), and do not suffer from the low-resource issue. Considering these characteristics of ASR task, we have raised a hypothesis: is it possible to improve hypernasality estimation by leveraging the advantages of the ASR objective.


As aforementioned, hypernasality derives abnormal expressions in the acoustic space (\eg, vowel and consonant).  The existing advanced methods for hypernasality estimation are mainly using neural network~\cite{Vikram2021Attention,Xiyue2019HypernasalityNet,Vikram2021Deep,Wang2019cnn}. But all of them just simply stacked multiple neural network layers over the acoustic features (like MFCCs) to conduct classification, and cannot enable model to further extract better acoustic semantics since these methods only focused on learning hypernasality-related features and ignored to understand acoustic semantics. However, it is worthy to note that ASR models can naturally extract high-quality representations of acoustic features (especially in identifying vowel and consonant), since large-scale text annotations encourage model to thoroughly understand the content of acoustic features (\eg, for text ``am" and its phonation /\textschwa m/, the model needs to identify vowel /\textschwa/ and consonant /m/, and then returns the answer). Inspired by success of transfer learning in other tasks~\cite{Huh2016imagenet,Jacob2019BERT}, we deem that using ASR model for initialization is beneficial for estimating hypernasality.

Therefore, in this paper, we introduce a novel approach to leverage ASR model for hypernasality estimation from the perspective of transfer learning. Specifically, we first pre-train an encoder-decoder framework by using the ASR objective on large-scale ASR corpus, and then apply the pre-trained ASR encoder to conduct hypernasality estimation on CLP corpus. Such design can take advantages of ASR model in learning hypernasality-related semantics: 1) ASR tasks usually includes enormous audio data that enables model to obtain a better generalization; 2) Benefiting from text annotations, the neural network based encoder with ASR objective is able to extract better phonetic representations for hypernasality estimation. Experimental results on two different CLP datasets also indicate that our model with ASR initialization is superior to the base model without using initialization. 



\begin{figure}[!t]
    \centering
    \includegraphics[width=0.45\textwidth]{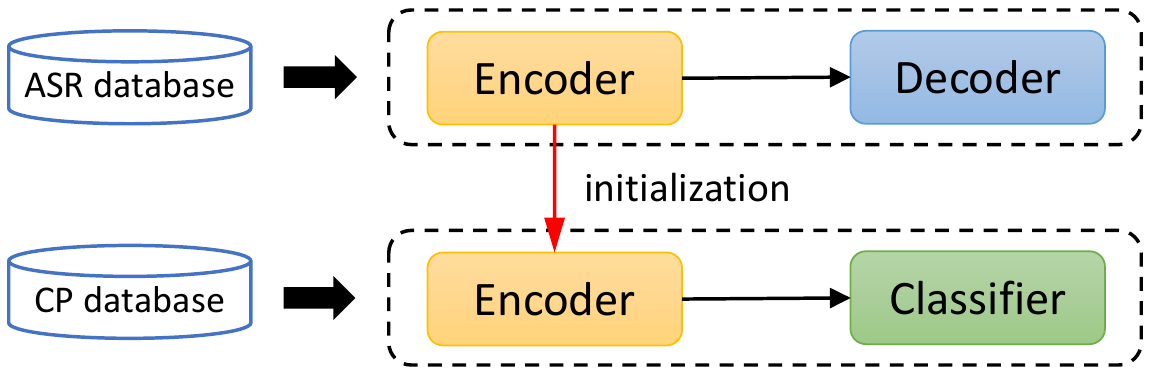}
    \caption{The training pipeline of our approach. The first row is using ASR dataset for model training, and then use ASR-trained encoder for an initialization to conduct hypernasality assessment on cleft palate dataset.}
    \label{fig_asr}
\end{figure}

\section{Method}
\label{sec2:method}

\subsection{ASR Formulation}
Automatic speech recognition (ASR) is to identify the content of human voices and then return the corresponding text. Assuming ASR corpus as ($\mathcal{S}, \mathcal{T}$), where $\mathcal{S}$ and $\mathcal{T}$ correspond to audios and the corresponding language descriptions, respectively. For audio data, they are first processed as a sequence of acoustic features (\eg, Mel filter-bank features) at the frame level, and then feed into the model. Generally, we adopt an encoder-decoder framework~\cite{Sutskever2014sequence} to handle ASR task where the decoder is formulated as an auto-regressive manner, and the parameters are defined as $\theta = \{\theta_{enc}, \theta_{dec}\}$. The objective of ASR system is to predict the most possible corresponding text from the given speech as $\arg \max P(\cal{T}|\cal{S})$. The objective function is optimized by maximum likelihood estimation as:
\begin{equation}
    \label{eq1}
    \mathcal{L}_{ASR} = - \sum_{(s, t)}^{(\mathcal{S}, \mathcal{T})} \sum_{i=1}^{|t|} \log P(t_i | t_{<i}, s; \theta_{enc}, \theta_{dec}),
\end{equation}
where $(s, t)$ is a paired speech and text, $t_{<i}$ means tokens before position $i$ and $\{\theta_{enc}, \theta_{dec}\}$ represent the parameters of encoder and decoder, and $P(t_i | t_{<i}, s; \theta_{enc}, \theta_{dec})$ is to predict the $i$-th token in the text sequence based on previous position and the given speech. Therefore, by optimizing the Eqn.~\ref{eq1}, we can obtain a well-trained ASR model in the general domain, and its encoder can extract deep representations from acoustic features so that the decoder is able to predict the corresponding text accurately. 




\subsection{Hypernasality Estimation Formulation}

The target of hypernasality estimation is to evaluate the level of hypernasality from the speech of CLP patients. More specifically, by adopting different evaluation criteria, hypernasality estimation can be subdivided as hypernasality detection and hypernasality assessment. The former is to detect whether the cleft palate speech  includes hypernasality or not, which can be simplified as a binary classification task. The latter is to predict the hypernasality rating of the corresponding cleft palate speech to determine the hypernasality severity, which is a more challenging multi-class classification task. In this paper, the rating of hypernasality severity is formulated as a range of 0$-$3 (0-normal, 1-mild, 2-moderate, 3-severe), each of which represents one class in the multi-class classification task. 

The CLP corpus can be denoted as a 2-tuple ($\mathcal{X}, \mathcal{Y}$), where $\mathcal{X}$ and $\mathcal{Y}$ represent the datasets of CLP speech and the corresponding hypernasality rating, respectively. For CLP speech, we adopt the same preprocessing step to obtain the similar acoustic features  (\ie, Mel filter-bank feature), like ASR task, to guarantee the consistency. To conduct hypernasality estimation, we employ an encoder network with a classifier layer, where the parameters are defined as $\tilde{\theta} = \{\tilde{\theta}_{enc}, \tilde{\theta}_{cls}\}$. Therefore, the objective of hypernasality estimation is formulated as following:
\begin{equation}
    \label{eq2}
    \mathcal{L}_{CLS} = - \sum_{(x, y)}^{(\mathcal{X}, \mathcal{Y})} \log P(y |x; \tilde{\theta}_{enc}, \tilde{\theta}_{cls}),
\end{equation}
where $\tilde{\theta}_{enc}$ and $\tilde{\theta}_{cls}$ represent the parameters of the encoder and classifier respectively.  The classifier is a linear layer to scale the dimension of the encoder outputs as the number of categories $|C|$, with a connected softmax function to predict probability. $|C|$ is respectively set as 2 and 4 for hypernasality detection and hypernasality assessment.

\subsection{Transfer Learning}
As mentioned above, with the help of large-scale speech data and the labeled text, the ASR encoder can be viewed as a powerful acoustic feature extractor. Consequently, we deem that the encoder with ASR training is better for hypernasality estimation due to its ability in learning high quality acoustic representation. To fulfill this target, we require the architecture of encoder used in the ASR task and hypernasality estimation  to be completely identical, and same as the input acoustic features (Mel filter-bank). More specifically, we first train an encoder-decoder framework in an ASR objective, and then conduct hypernasality estimation task based on ASR encoder. Therefore, the objective function of our method is as follows: $$\mathcal{L}_{CLS} = - \sum_{(x, y)}^{(\mathcal{X}, \mathcal{Y})} \log P(y |x; \theta_{enc}, \tilde{\theta}_{cls}),$$ where $\theta_{enc}$ is the parameters of ASR encoder in Eqn.~\ref{eq1}. Figure~\ref{fig_asr} also presents a simple description about our training pipeline.

\section{Experiments}
\label{sec3:exp}

\begin{table}[!t]
    \centering
    \begin{tabular}{l|r|r}
    \toprule
    Rating  &  \# NMCPC & \# CNH \\
    \midrule 
    Normal &  25 & 239 \\
    Mild   &  11 & 190 \\
    Moderate & 14 & 509 \\
    Severe & 16 & 108 \\
    \bottomrule
    \end{tabular}
    \caption{Statistics of NMCPC and CNH cleft palate datasets. `` \# NMCPC'' and ``\# CNH" column represent the patient number of each hypernasality severity.}
    \label{tab:my_label}
\end{table}

\subsection{CLP dataset}
\subsubsection{NMCPC-CLP}
NMCPC-CLP is a dataset collected by New Mexico Cleft Palate Center~\cite{Mohammad2020Single}, which is mainly composed of English speakers with cleft palate. This dataset includes 41 CLP patients and we sample 25 normal speakers as the control group. The average age of NMCPC-CLP is 9.2$\pm$3.3 years. Each patient is required to record a random subset of sentences from the candidate sentences. Based on the sampled audios, each patient will be assigned a score from 0 to 3 (0 stands normal and 3 means severe). For hypernasality detection, the patients with hypernasality are defined as who has a score between 1 and 3. Refer to Table~\ref{tab:my_label} for more details about the NMCPC-CLP dataset.

\subsubsection{CNH-CLP}
CNH-CLP is a dataset collected by a Chinese hospital, which includes cleft palate patients with a range from children to adults. All patients in CNH-CLP dataset are Chinese speakers. The procedure of audio collection for each patient is similar as NMCPC-CLP dataset. Table~\ref{tab:my_label} introduces the detailed information about the CNH dataset.

\subsection{ASR Datasets}
\subsubsection{Librispeech} 
Librispeech~\cite{Vassil2015Librispeech} is a large-scale speech recognition dataset in English domain. Librispeech includes 960 hrs speech data, sampled at 16,000 Hz with the corresponding text for training. During the ASR training, we select dev-clean/dev-other as the development set. We use Librispeech to train an ASR model, and then apply it to NMCPC-CLP dataset since these two datasets are both in  English domain.

\subsubsection{Aishell} 
Aishell-1~\cite{Hui2017Aishell} is a widely used speech recognition dataset in Chinese domain. Aishell-1 dataset includes 150/10 hrs audios for the training/dev set. Each audio is sampled at 16,000 Hz, with the corresponding text. We use Aishell-1 for ASR training and then use it for CNH-CLP dataset as these two tasks belong to the same language.

\begin{table}[t]
    \centering
    \begin{tabular}{l l |rr}
    \toprule
      & & ASR & CLS \\
    \midrule 
    \multirow{3}{*}{CNN} & stride       & \multicolumn{2}{c}{2, 2} \\
                         & kernel width & \multicolumn{2}{c}{5, 5} \\
                         & channel      & \multicolumn{2}{c}{1024} \\
    \midrule
    \multirow{5}{*}{Transformer} & layer & \multicolumn{2}{c}{12}  \\
                         & hidden size   & \multicolumn{2}{c}{512} \\
                         & filter size   & \multicolumn{2}{c}{2048}\\
                         & dropout       & \multicolumn{2}{c}{0.1} \\
                         & heads         & \multicolumn{2}{c}{8}   \\
    \midrule
    \multirow{3}{*}{Optimization}         & batch         & 256  & 32 \\
                         & learning rate & 2e-3 & 2e-4 \\
                         & Epoch         & 100  & 30 \\
    \bottomrule
    \end{tabular}
    \caption{Hyper-parameters of experimental setup. The ``ASR'' and ``CLS'' column means the setting of automatic speech recognition, and hypernasality estimation.}
    \label{tab:hyper}
\end{table}

\subsection{Setup}
For audio inputs, we first resample audios at 16,000 Hz and then extract 80-channel log mel filter-bank features (25 ms window size and 10 ms shift) for all ASR datasets and CLP datasets. We choose Fairseq-S2T~\cite{wang-etal-2020-fairseq} as the development toolkit. For ASR task, the encoder is composed of two convolutional layers for sub-sampling and a stack of transformer layers~\cite{Vaswani2017transformer}, and the decoder is a stack of transformer layers with cross-attention modules. During the ASR training stage, we use SpecAugment~\cite{Daniel2020SpectAug} for data augmentation. We adopt Adam~\cite{Kingma2015Adam} as the default optimizer. The detailed hyperparameters are reported in Table~\ref{tab:hyper}. In addition, to reduce variance, we use five-fold cross-validation to measure our classification accuracy. All evaluations are tested at the speaker level.

\section{Results}
\label{sec:results}

\subsection{Hypernasality Estimation Accuracy}
We adopt the precision as the metric to evaluate the performance of our model in hypernasality assessment. The results are shown in Table~\ref{tab:results}. Our baseline is the base model without using ASR encoder. We also list some baselines~\cite{Mohammad2020Single}, which use statistical features for reference. From Table~\ref{tab:results}, we have the following observations: 1) when configured with ASR encoder, our method can outperform the baseline method by a large margin, including hypernasality detection and assessment, in both NMCPC and CNH cleft palate datasets; 2) compared with the statistical methods, our method also achieves significant improvement, especially in hypernasality detection. These improvements also demonstrate the effectiveness of our method. Besides, our method is capable of generality and can be applied to any neural networks used in previous works~\cite{Dubey2019Hypernasality,Vikram2021Attention}.

\begin{table}[h]
    \centering
    \begin{tabular}{l|c c| c c}
    \toprule
             & \multicolumn{2}{c|}{NMCPC} & \multicolumn{2}{c}{CNH} \\
    Method   & \multicolumn{1}{c}{HD} & \multicolumn{1}{c|}{HA} & \multicolumn{1}{c}{HD} & \multicolumn{1}{c}{HA} \\
    \midrule 
    Baseline & {\scriptsize 90.1 $\pm$ 0.3}  & {\scriptsize 77.8 $\pm$ 0.6} & {\scriptsize 93.1 $\pm$ 0.3}    & {\scriptsize 73.1 $\pm$ 0.3} \\ 
    + ASR & {\scriptsize 93.4 $\pm$ 0.2} & {\scriptsize 83.6 $\pm$ 0.2} &   {\scriptsize 96.5 $\pm$ 0.2} & {\scriptsize 79.8 $\pm$ 0.5}  \\
    \midrule 
    MFCC~\cite{Mohammad2020Single}    & 84.07 & 62.64 & {\scriptsize 88.1 $\pm$ 0.2} & {\scriptsize 69.1 $\pm$ 0.1} \\
    CQCC~\cite{Mohammad2020Single}    & 84.07 & 70.05 & - & - \\
    SFFB~\cite{Mohammad2020Single}    & 89.00 & 82.10 & - & - \\
    \bottomrule
    \end{tabular}
    \caption{Results of hypernasality estimation on NMCPC and CNH cleft palate datasets. The ``+ ASR'' row means using ASR encoder. ``HD'' and ``HA" represent hypernasality detection and hypernasality assessment respectively.}
    \label{tab:results}
    \vspace{-0.3cm}
\end{table}

\subsection{Analysis}

\begin{figure}[h]
    \centering
    \includegraphics[width=0.4\textwidth]{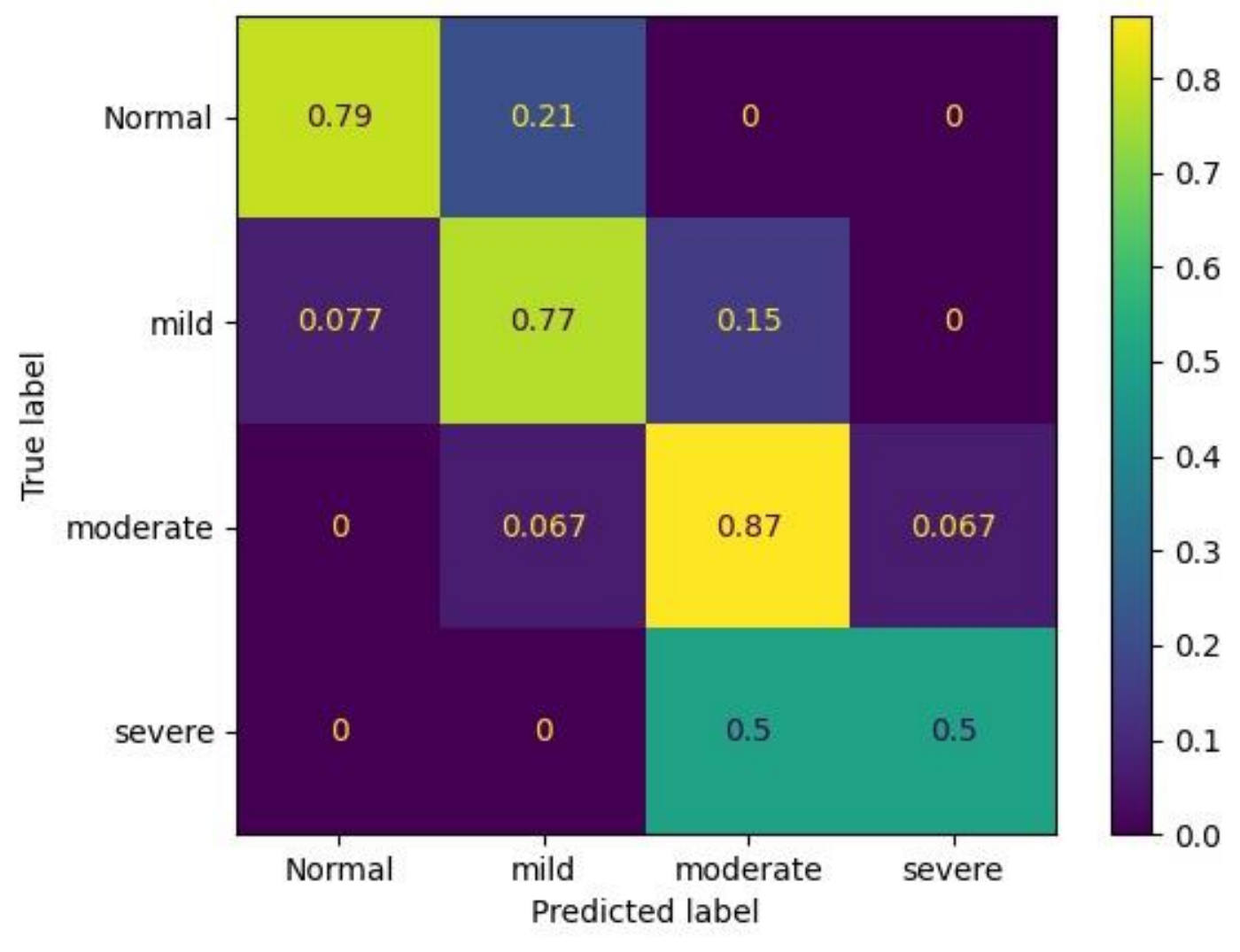}
    \caption{Confusion matrix of our method for hypernasality classification on CNH dataset.}
    \label{fig_cm}
\end{figure}

\subsubsection{Confusion Matrix}
To demonstrate the generalization of our model in predicting hypernasality, we also visualize the confusion matrix of our model by using ASR encoder in hypernasality classification, and the results are shown in Figure~\ref{fig_cm}. From Figure~\ref{fig_cm}, we observe that our model is significant in predicting each label, especially in normal, mild and moderate patients. Considering that CNH is an unbalanced dataset (moderate and severe patients occupy 50\% and 10\% respectively), the accuracy of our model in predicting severe cases is still acceptable. Overall, these results indicate that our model can obtain better generalization performance and avoid overfitting effectively.

\subsubsection{Visualization}
\begin{figure}[h]
    \centering
    \includegraphics[width=0.45\textwidth]{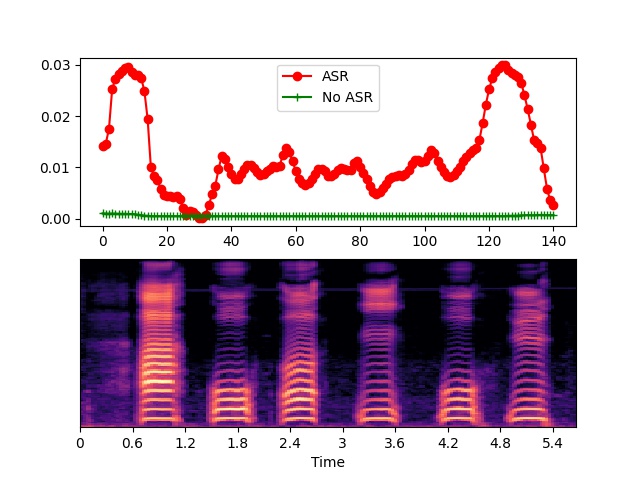}
    \caption{Comparisons between our method and baseline without using ASR in the activated value of encoder output.}
    \vspace{-0.2cm}
    \label{fig_attn}
    \vspace{-0.2cm}
\end{figure}

To better explain the advantages of our methods in identifying acoustic features, we also visualize the hidden unit of encoder output. More specifically, we assume the output of encoder as $h \in \mathbb{R}^{L \times D}$, where $L$ and $D$ represent the frame length and the hidden size, respectively. Therefore, we first calculate the average value of $h$, and then use its absolute value as the activated value of $i$-th frame (\ie, $\frac{1}{D} |\sum_{j=1}^D {h_{i,j}}|$). The result and its corresponding mel spectrum are shown in Figure~\ref{fig_attn}. We can find that the outputs of our model is more sensitive to acoustic features and demonstrate higher activated value, while the model without using ASR is too smooth to distinguish the semantic of acoustic features. This phenomenon also validates our hypothesis that ASR model can indeed help to extract high quality acoustic features.

\section{Conclusion}
In this paper, we introduce a simple and effective approach to improve hypernasality estimation from the perspective of using ASR model. More specifically, we note that existing neural network based methods only stack multiple neural network layers for classification, and cannot extract high quality representation for acoustic features, which is useful for hypernasality estimation. To address this deficiency, we propose to fine tune the encoder, which is pre-trained with the ASR objective, for hypernasality estimation. Such design allows our model to enjoy the benefits of ASR model from two aspects: 1) ASR corpus usually includes more audio data, which enables better generalization; 2) the labeling text of ASR corpus guides model to better extract acoustic features. Experimental results on two cleft palate datasets also demonstrate the effective of our methods in hypernasality assessment. In the future, we expect to focus on two research directions: 1) is it possible to utilize more powerful pre-trained speech models (like wav2vec~\cite{Steffen2019wav2vec}) to conduct hypernasality estimation in cleft palate speech; 2) traditional methods for hypernasality estimation usually designed some advanced statistical acoustic features, and thus we want to explore the potential of combining neural network based features and statistical features to better estimate hypernasality.

\label{sec:conclusion}


\bibliographystyle{IEEEtran}

\bibliography{mybib}


\end{document}